\documentclass[english,floatfix,twocolumn,superscriptaddress,prb,aps]{revtex4}
\usepackage[T1]{fontenc}
\usepackage[latin1]{inputenc}
\usepackage{graphicx}
\usepackage{amssymb}
\usepackage{amsmath}
\usepackage{bm}
\usepackage{babel}
\usepackage{color}
\makeatother

\renewcommand{\Re}{\mathop{\mathrm{Re}}\nolimits}
\renewcommand{\Im}{\mathop{\mathrm{Im}}\nolimits}
\renewcommand{\vec}[1]{\mathbf{#1}}

\begin{document}
\title{Electron-electron interactions and doping dependence of the two-phonon Raman intensity in graphene}
\author{D.~M.~Basko}\email{denis.basko@grenoble.cnrs.fr}
\affiliation{Laboratoire de Physique et Mod\'elisation des Mileux Condens\'es,
Universit\'e Joseph Fourier and CNRS, Grenoble, France}
\author{S. Piscanec}
\author{A.~C. Ferrari}
\affiliation{Department of Engineering, Cambridge University,
9 JJ Thomson Avenue, Cambridge CB3 OFA, UK}

\begin{abstract}
Raman spectroscopy is a fast, non-destructive means to characterize graphene samples. In particular, the Raman spectra are strongly affected by doping. While the change in position and width of the G peak can be explained by the non-adiabatic Kohn anomaly at $\Gamma$, the significant doping dependence of the 2D peak intensity has not been explained yet. Here we show that this is due to a combination of electron-phonon and electron-electron scattering. Under full resonance, the photogenerated electron-hole pairs can scatter not just with phonons, but also with doping-induced electrons or holes, and this changes the intensity. We explain the doping dependence and show how it can be used to determine the corresponding electron-phonon coupling. This is higher than predicted by density-functional theory, as a consequence of renormalization by Coulomb interactions.
\end{abstract}


\maketitle

\section{Introduction}
Graphene is the latest carbon allotrope to be discovered, and it is now at
the center of a significant research effort\cite{Nov306(2004),GeimRevNM, Nov438(2005),CastroNetoRev,charlier,Zhang438(2005)}. Near-ballistic transport at room temperature and high mobility\cite{Nov438(2005),Zhang438(2005),Nov315(2007),MorozovNov(2007),andrei,kimmob} make it a
potential material for nanoelectronics\cite{Han, Chen,Zhang86,Lemme}, especially for high frequency
applications\cite{Yuming}. Furthermore, its transparency and mechanical properties are ideal for micro and nanomechanical systems, thin-film transistors and transparent and conductive composites and electrodes\cite{bunch,blake1,hernandez,eda}.

\begin{figure}
\includegraphics[width=85mm]{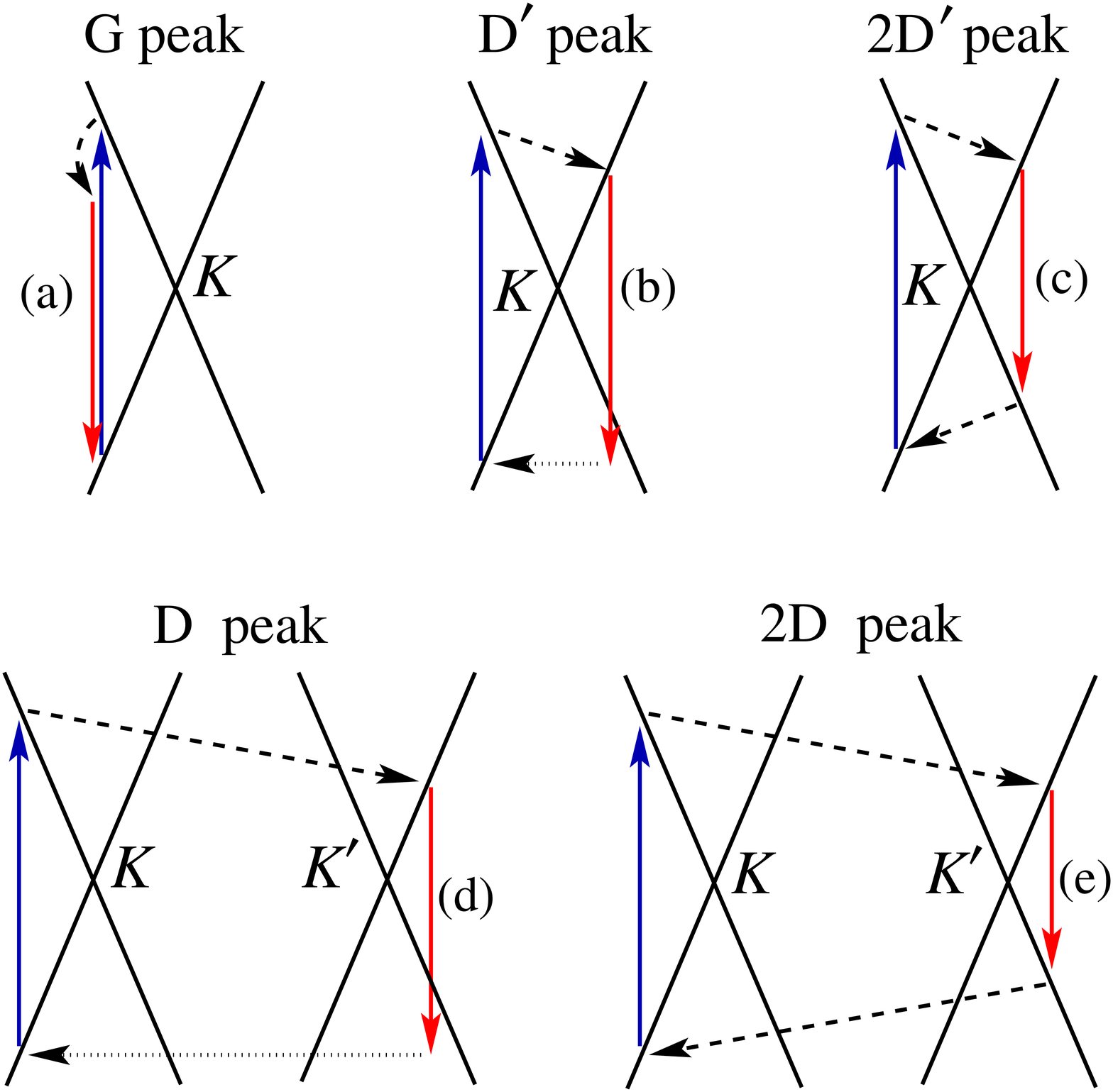}
\caption{(Color Online) Role of the electron dispersion
(Dirac cones, $\epsilon=\pm v_F|\vec{p}|$, shown by solid black lines) in Raman scattering:
(a)~intravalley one-phonon $G$~peak, (b)~defect-assisted intravalley one-phonon $D'$~peak,
(c)~intravalley two-phonon $2D'$~peak, (d)~defect-assisted intervalley one-phonon $D$~peak,
(e)~intervalley two-phonon $2D$~peak. Vertical solid arrows represent interband transitions accompanied
by photon absorption (blue lines) or emission (red lines) (the photon wavevector is neglected). Dashed
arrows represent phonon emission. Horizontal dotted arrows represent defect scattering.}
\label{fig:res}
\end{figure}

Graphene layers can be readily identified in terms of number and orientation by inelastic and
elastic light scattering, such as Raman\cite{ACFRaman} and Rayleigh
spectroscopies\cite{CasiraghiNL,GeimAPL}. Raman spectroscopy also allows monitoring of doping,
defects, strain, disorder, chemical modifications and edges\cite{ACFRaman,CasiraghiAPL,Pisana,ACFRamanSSC,DasCM,Mohi,ferralis,YanPrl2007,Leandro,PiscanecPRL,cancado08,Cedge,Elias,Dasbila,Ferrari00,Graf2007}.
Indeed, Raman spectroscopy is a fast and non-destructive characterization method for carbons\cite{acftrans}. They
show common features in the 800-2000 cm$^{-1}$ region: the G and D peaks, around 1580 and 1350
cm$^{-1}$, respectively. The G peak corresponds to the $E_{2g}$ phonon at the Brillouin zone center
($\bf\Gamma$~point). The D peak is due to the breathing modes of six-atom rings and requires a defect for its
activation\cite{tuinstra,Ferrari00,ThomsenPrl2000}. It comes from TO phonons around the \textbf{K} point of the
Brillouin zone\cite{tuinstra,Ferrari00}, is active by double resonance (DR)\cite{ThomsenPrl2000}, and is
strongly dispersive with excitation energy due to a Kohn Anomaly at \textbf{K}\cite{PiscanecPRL}.
The activation process for the D peak is inter-valley, and is shown
schematically in Fig.~\ref{fig:res}(d): i) a laser induced
excitation of an electron/hole pair; ii) electron-phonon scattering with an exchanged momentum
$\textbf{q}\sim\textbf{K}$; iii) defect scattering; iv) electron-hole recombination.
DR can also happen as intra-valley process, i.~e. connecting two points belonging to
the same cone around $\textbf{K}$ (or $\textbf{K}'$), as shown in Fig.~\ref{fig:res}(b).
This gives the so-called D'peak, which is at$\sim1620~cm^{-1}$ in defected graphite measured at 514nm.

The 2D peak is the second order of the D peak. This is a single peak in single layer graphene (SLG), whereas it splits in four in bilayer graphene (BLG), reflecting the evolution of the band structure\cite{ACFRaman}. The 2D' peak is the second order of the D' peak. Since both 2D and 2D' originate from a process where
momentum conservation is satisfied by two phonons with opposite wavevectors ($\textbf{q}$ and $-\textbf{q}$), they do not require the presence of defects for their activation, and are thus always present. Indeed, high quality graphene shows the G, 2D and 2D' peaks, but not D
and D'\cite{ACFRaman}. Also, under the assumption of electron-hole symmetry, the two-phonon peaks are fully resonant\cite{Basko2007,BaskoBig}.
This means that energy and momentum conservation are satisfied in all elementary steps of the Raman
process, as shown schematically in Fig.~\ref{fig:res}(c,e). Then, all intermediate electronic states are real. As a consequence, two-phonon Raman spectroscopy is sensitive to the dynamics of the photo-excited electron-hole pair, in particular, to the scattering processes it can undergo.
This is of crucial importance for the present work.

The effects of doping on the graphene G-peak position [Pos(G)] and Full Width at Half Maximum [FWHM(G)] were reported in Refs. \onlinecite{Pisana,YanPrl2007,DasCM,Dasbila}. Pos(G) increases and FWHM(G) decreases for both electron and hole doping. The G peak stiffening is due to
the non-adiabatic removal of the Kohn-anomaly at $\bf{\Gamma}$\cite{Lazzeri2006,Pisana}. The FWHM(G) sharpening is due to Pauli blocking of
phonon decay into electron-hole pairs, when the electron-hole gap is higher than the phonon energy\cite{LazPRB2006,Pisana}, and saturates for
a Fermi shift bigger than half phonon energy\cite{Pisana,YanPrl2007,LazPRB2006}. A similar behavior is observed for the LO-G$^{-}$ peak in metallic nanotubes\cite{dasnt}, for the same reasons. In the case of BLG, the different band structure re-normalizes the phonon
response to doping differently from SLG\cite{Yan2008,Dasbila,andobi}. Also in this case the Raman
G peak stiffens and sharpens for both electron and hole doping, as a result of the
non-adiabatic Kohn anomaly at $\bf\Gamma$\cite{Dasbila}. However, since BLG has two conduction and valence
subbands, with splitting dependent on the interlayer coupling, this changes the slope in the variation of Pos(G) with doping, allowing a direct
measurement of the interlayer coupling strength\cite{Dasbila,andobi}.

\begin{figure}
\includegraphics[width=70mm]{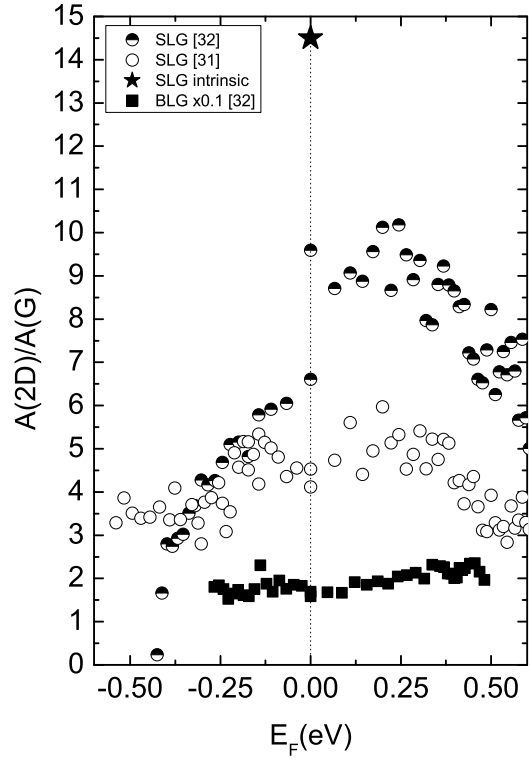}
\caption{Experimental A(2D)/A(G), measured for 514.5nm excitation, as a function of $E_F$ for SLG\cite{Dasbila,DasCM, ACFRaman,heinz,berciaud} and BLG\cite{Dasbila}. The BLG data (solid squares) are divided by 10, to make comparison easier. Note that the doping dependent SLG data are a combination of two experiments on two different samples, from Ref.\onlinecite{Dasbila} (half-filled circles) and Ref.\onlinecite{DasCM} (Open circles), and a data-point representative of intrinsic graphene from Refs.\onlinecite{ACFRaman,heinz,berciaud,tan1} (solid star)}
\label{figdata}
\end{figure}

Another significant result is that in SLG the ratio of the heights of the 2D and G peaks, I(2D)/I(G), and their areas, A(2D)/A(G), is maximum for zero doping\cite{ACFRaman,heinz,berciaud}, and decreases for increasing doping. On the other hand, this shows little dependence on doping for BLG\cite{DasCM,Dasbila}. Fig.~\ref{figdata} plots the combined data for SLG and BLG from Refs.~\onlinecite{DasCM,Dasbila,ACFRaman,heinz,berciaud,tan1}. Note that Refs.~\onlinecite{DasCM,Dasbila} reported height ratios, while here, as discussed later, we analyze the area ratio A(2D)/A(G), which encompasses both trends of I(2D)/I(G) and FWHM(2D)/FWHM(G).

Due to residual disorder, the energy of the Dirac point can fluctuate across the sample on a scale smaller than the laser spot, which leads to spatial inhomogeneity of the doping level\cite{CasiraghiAPL,yacoby}. We attribute the difference in the behavior of the two SLG curves in Fig.2 to a different degree of residual charge inhomogeneity in the polymeric electrolyte experiments of Refs.~\onlinecite{DasCM,Dasbila}. On the other hand, the use of this electrolyte enabled probing a very large doping range, because the nanometer-thick Debye layer gives a much higher gate capacitance compared to the usual 300nm SiO$_2$ back gate~\cite{DasCM,Dasbila,Pisana}. Note as well that A(2D)/A(G) for the most intrinsic samples measured to date is $\sim$12--17~\cite{ACFRaman,heinz,berciaud,tan1}, much higher than the zero gating values in Refs.~\onlinecite{DasCM,Dasbila}, as shown in Fig. 2. This points again to sources of disorder in the gated samples of Refs.~\onlinecite{DasCM,Dasbila}, while the absence of a significant D peak excludes large amounts of structural defects.

Here, we show that the 2D intensity doping dependence results from its sensitivity to the scattering of the photoexcited electron and hole. Assuming the dominant sources of scattering to be phonon emission and electron-electron collisions, we note that, while the former is not sensitive to doping, the latter is. Then, the 2D doping dependence can be used to estimate the corresponding electron-phonon coupling (EPC).
\section{Doping Dependence of Two Phonon Raman Intensity}
\subsection{Theoretical Dependence}
Raman scattering\cite{Raman} is an electron-mediated process where electromagnetic radiation exchanges vibrational quanta (phonons) with a crystal. A complete description requires the detailed knowledge of (i) electronic structure, (ii) phonon dispersions, (iii) mutual interactions between electrons and phonons (i.e. electron-electron, electron-phonon and phonon-phonon scattering).

The Raman spectrum of graphene consists of a set of distinct peaks. Each characterized by its position width, height, and area. The frequency-integrated area under each peak represents the probability of the whole process. It is more robust with respect to various perturbations of the phonon states than width and height. Indeed, for an ideal case of dispersionless undamped phonons with frequency $\omega_\mathrm{ph}$ the shape of the $n$-phonon peak is a Dirac $\delta$ distribution $\propto\delta(\omega-n\omega_\mathrm{ph})$, with zero width, infinite height, but well-defined area. If the phonons decay (e.~g, into other phonons, due to anharmonicity, or into electron-hole pairs, due to electron-phonon coupling), the $\delta$ lineshape broadens into a Lorentzian, but the area is preserved, as the total number of phonon states cannot be changed by such perturbations. If phonons have a weak dispersion, states with different momenta contribute at slightly different frequencies. This may result in an overall shift and a non-trivial peak shape, but frequency integration across the peak means counting all phonon states, as in the dispersionless case. Thus, the peak area is preserved, as long as the Raman matrix element itself is not changed significantly by the perturbation. The latter holds when the perturbation (phonon broadening or dispersion) is smaller than the typical energy scale determining the matrix element. Converting this into a time scale using the uncertainty principle we have that, if the Raman process is faster than the phonon decay, the total number of photons emitted within a given peak (i.~e., integrated over frequency across the peak), is not affected by phonon decay, although their spectral distribution can be. Although the graphene phonons giving rise to the D and D' peak are dispersive due to the Kohn Anomalies at $\textbf{K}$ and $\bf{\Gamma}$ \cite{PiscanecPRL}, their relative change with respect to the average phonon energy is at most a few \%, thus we are in the weakly dispersive case discussed above. The phonon decay in graphene is in the picosecond timescale, while the Raman process is faster, in the femtosecond timescale\cite{Pisana,Bonini2007,LazzeriHP}. Then, we will analyze the area ratio, A(2D)/A(G), which encompasses both variations in height ratio, I(2D)/I(G), and width: FWHM(2D)/FWHM(G).

We first consider the G peak. For the one-phonon process, allowed by momentum conservation, which gives rise to the G peak, the picture is entirely different from the two-phonon case. As shown in Fig.1a, the process responsible for the G peak is determined by virtual electron-hole pairs with energy $E_L/2$, where $E_L$ is the laser excitation energy (for a typical visible Raman measurement $E_L/2\sim1\:\mbox{eV}$). If the Fermi energy, $E_F$, stays below $E_L/2$, as in Refs.~\onlinecite{DasCM,Dasbila}, these electronic states are not strongly affected. Only the final phonon state is influenced by doping, which manifests itself in a change of Pos(G) and FWHM(G)\cite{DasCM,Dasbila,Pisana,YanPrl2007}. However, the area of the peak is determined by the total spectral weight of the phonon state, which is preserved. Thus, we do not expect any significant dependence of A(G) on doping, as long as the doping is not too strong, so that $|E_F|\ll{1}\:\mbox{eV}$. We can then take the measured doping dependence of A(2D)/A(G) as representative of the A(2D) trend. Note that A(G) can change as a function of other external parameters, such as the Raman excitation energy\cite{Vidano, Pocsik1998,Cancado2007,ACFRaman,Ferrari00}. However, for fixed excitation, such as in the experiments discussed here, the above argument holds.

In Ref.\onlinecite{BaskoBig} the following expressions for the 2D and 2D' areas were obtained:
\begin{subequations}
\begin{eqnarray}\label{I2D=}
&&A(2D)=\frac{8}{3}\left(\frac{e^2}c\right)^2\frac{v_F^2}{c^2}
\left(\frac{\gamma_K}{\gamma}\right)^2,\\
&&A(2D')=\frac{4}{3}\left(\frac{e^2}c\right)^2\frac{v_F^2}{c^2}
\left(\frac{\gamma_\Gamma}{\gamma}\right)^2.\label{I2Dp=}
\end{eqnarray}\end{subequations}
where $e$ is the electron charge, $c$~is the speed of light, $e^2/c\approx{1}/137$ is the fine structure constant, and $v_F$ is the electron velocity (its experimental value is $v_F\approx{10}^6\:\mbox{m/s}\approx{6}.6\:\mbox{eV}\cdot\mbox{\AA}$\cite{Jiang2007,Rotenberg2007SSC,Zhou2008}). $2\gamma$~is the scattering rate of the photoexcited electron and hole. Note that we define $\gamma$~as the imaginary part of the energy, so it determines the decay of the amplitude, while the decay of the probability is determined by $2\gamma$. This includes all sources of inelastic scattering. Assuming the two main mechanisms for electron scattering to be the emission of phonons and electron-electron collisions, we write:
\begin{equation}
\gamma=\gamma_\mathrm{e-ph}+\gamma_\mathrm{ee},\quad
\gamma_\mathrm{e-ph}=\gamma_\Gamma+\gamma_K.
\end{equation}
Here we include the phonons near $\bf\Gamma$ and \textbf{K}, responsible for D and D'. The corresponding emission rates, $2\gamma_\Gamma$ and $2\gamma_K$, enter the numerators in Eqs.~(\ref{I2D=}), (\ref{I2Dp=}).

Two points regarding Eqs.~(\ref{I2D=}), (\ref{I2Dp=}) should be emphasized. First, the scattering rates depend on the electron energy,~$\epsilon$, which is defined by half the laser energy, $\epsilon\approx{E}_L/2$ [see Eq.~(\ref{gamma1=}) in the next section]. Second, if impurity scattering is significant compared to other scattering mechanisms, the corresponding elastic scattering rate cannot be simply included in~$\gamma$ and Eqs.~(\ref{I2D=}), (\ref{I2Dp=}). The whole Raman intensity calculation should be done differently. Eqs.~(\ref{I2D=}), (\ref{I2Dp=}) thus neglect impurity scattering. For short-range impurities this assumption is justified by the absence of a large D peak in the spectra of Refs.~\onlinecite{DasCM,Dasbila}. Long-range disorder is efficiently screened (even though the vanishing density of states at the Dirac point requires the screening to be nonlinear\cite{Fogler,Novikov,DasSarma,Polini}); it is precisely this screening that gives rise to the inhomogeneous concentration of electrons/holes and spatial fluctuations of the Dirac point energy.

In principle, there are no reasons for a strong dependence of~$\gamma_\mathrm{e-ph}$ on carrier density. However, $\gamma_\mathrm{ee}$ does exhibit such a dependence. Indeed, in undoped graphene at low temperatures, the photoexcited electron finds itself in a state with some momentum, $\vec{p}$, measured from the Dirac point, in the empty conduction band. To scatter into a state with a different momentum $\vec{p}'$, it has to give away some energy and momentum to another electron in the full valence band. This second electron would have to be promoted to the conduction band (as there are no available empty states in the valence band) into a state with momentum $\vec{p}_e$, leaving a hole in the valence band with ~$\vec{p}_h$. Momentum and energy conservation require:
\begin{subequations}\begin{eqnarray}
&&\vec{p}=\vec{p}'+\vec{p}_e+\vec{p}_h,\label{momcons=}\\
&&\epsilon(\vec{p})=\epsilon(\vec{p}')+\epsilon(\vec{p}_e)+\epsilon(\vec{p}_h),
\label{encons=}
\end{eqnarray}\end{subequations}
where $\epsilon(\vec{p})$ is the quasiparticle dispersion, assumed the same for electrons and holes. For Dirac particles, $\epsilon(\vec{p})=v_F|\vec{p}|$, the only possibility to satisfy both conservation laws is to have all four momenta parallel. If the spectrum is convex, $d^2\epsilon(p)/dp^2>0$, the two equations can be satisfied by a set of momenta with non-zero measure, i.~e. the phase space is finite. If it is concave, $d^2\epsilon(p)/dp^2<0$, they are incompatible. In SLG the spectrum is Dirac to a first approximation, resulting in an uncertainty\cite{Guinea1996}. This can be resolved by taking into account corrections from electron-electron interactions, which make the spectrum concave,\cite{AbrikosovBeneslavskii,Guinea94} and the interband process forbidden.

As new carriers are added to the system, intraband electron-electron collisions become allowed. The momentum and energy conservation become:
\begin{eqnarray}
&&\vec{p}+\vec{p}_e=\vec{p}'+\vec{p}_e',\\
&&\epsilon(\vec{p})+\epsilon(\vec{p}_e)=\epsilon(\vec{p}')+\epsilon(\vec{p}_e'),
\end{eqnarray}
which can be satisfied for any quasiparticle dispersion.
These collisions give a contribution to $\gamma_\mathrm{ee}$ which increases with carrier concentration. As a consequence, the total~$\gamma$ in Eq.1a increases, leading to an overall decrease of A(2D), consistent with the experimental trend in Fig.~\ref{figdata}.

The above arguments essentially use the non-convexity of the electronic spectrum in the conduction band, and thus apply to SLG only. In BLG, the spectrum is parabolic near the Dirac point, so that $d^2\epsilon/dp^2>0$, and the phase-space restrictions are absent. Thus, electron-electron collisions are allowed even at zero doping, and the collision rate has a much weaker dependence on $E_F$, which, in first approximation, can be neglected. Thus, A(2D) is expected to have a weak
dependence on~$E_F$, as seen in Fig.~\ref{figdata}, where the experimental A(2D)/A(G) for BLG shows a negligible variation with doping\cite{Dasbila}.

\begin{figure}
\includegraphics[width=80mm]{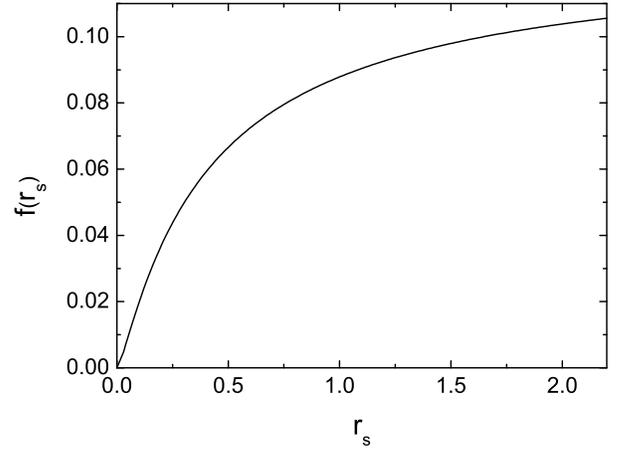}
\caption{Numerical values of $f(r_s)$, from Eq.(\ref{eerate=})}
\label{fig:plot}
\end{figure}

To quantify the doping effects on the SLG A(2D), we first calculate the electron-electron scattering rate, $2\gamma_\mathrm{ee}$, in the random-phase approximation, analogously to Refs.~\onlinecite{DasSarma2008,Polini2008}. $\gamma_\mathrm{ee}$ is given by the imaginary part of the on-shell electronic self-energy, $\Im\Sigma_{ee}(p,\epsilon)$ for $\epsilon\to{v_F}p-0^+$, with $\epsilon$ and \textbf{p} counted from the Dirac point\cite{Guinea1996}. Here we consider the limiting case, when the energy of the photoexcited electron ($\epsilon=E_L/2$) far exceeds $E_F$. The carrier concentration is $n=E_F^2/(\pi{v_F}^2)$. In this case, the collisions are dominated by small momentum transfers, $|\vec{p}-\vec{p}'|\sim|E_F|/v_F$, so $\gamma_{\mathrm{ee}}$ does not depend on~$\epsilon$ and is proportional to~$|E_F|$, the proportionality coefficient depending only on the dimensionless Coulomb coupling constant $r_s=e^2/(\varepsilon{v_F})$ ($\varepsilon$ being the dielectric constant):\begin{equation}
\label{eerate=}
\gamma_\mathrm{ee}=|E_F|\,f\!\left(\frac{e^2}{\varepsilon{v}_F}\right)
+O(E_F^2/\epsilon),
\end{equation}
where the function $f$ is given by:
\begin{eqnarray}
&&\label{f=}
f(r_s)=\frac{2}\pi\int\limits_0^{\pi/2}{d}\varphi \nonumber\\
&&\times \left\{\int\limits_0^{2/(1+\cos\varphi)}\frac{dx\;x^2\sin\varphi\,R_1}
{[2(x/r_s+4)x\sin\varphi]^2+R_1^2}\right. \nonumber\\
&&+\left.\int\limits_{2/(1+\cos\varphi)}^{2/(1-\cos\varphi)}
\frac{dx\;x^2\sin\varphi\,R_2}{[2(x/r_s+4)x\sin\varphi-R_3]^2
+R_2^2(x,\varphi)}\right\},\nonumber\\
\end{eqnarray}
and $R_1$, $R_2$, $R_3$ are:
\begin{subequations}\begin{eqnarray}
&&R_1(x,\varphi)=a_+b_+-a_-b_--x^2\ln\frac{a_++b_+}{a_-+b_-},\\
&&R_2(x,\varphi)=a_+b_+-x^2\ln\frac{a_++b_+}{x},\\
&&R_3(x,\varphi)=a_-\sqrt{x^2-a_-^2}-x^2\arccos\frac{a_-}x,\\
&&a_\pm=2\pm{x}\cos\varphi,\quad{b}_\pm=\sqrt{a_\pm^2-x^2}.
\end{eqnarray}\label{R=}
\end{subequations}
Fig.~\ref{fig:plot} plots $f(r_s)$, calculated numerically.

Thus, we expect A(2D) to change with $E_F$ as:
\begin{equation}\label{fitexpr=}
A(2D)=\frac{C}{[\gamma_{e-ph}+|E_F|f(e^2/\varepsilon{v}_F)]^2}
\end{equation}
with $C$ a constant. Note that a variation of the dielectric constant $\varepsilon$ will affect A(2D). Given the negligible dependence of A(G) on doping, Eq.~(\ref{fitexpr=}) can be rewritten as
\begin{equation}\label{fitexpr2=}
\sqrt{\frac{A(G)}{A(2D)}}=C'[\gamma_{e-ph}+|E_F|f(e^2/\varepsilon{v}_F)],
\end{equation}
where $C'$ is another constant.
\subsection{Fit to Experiments}
\begin{figure}
\includegraphics[width=80mm]{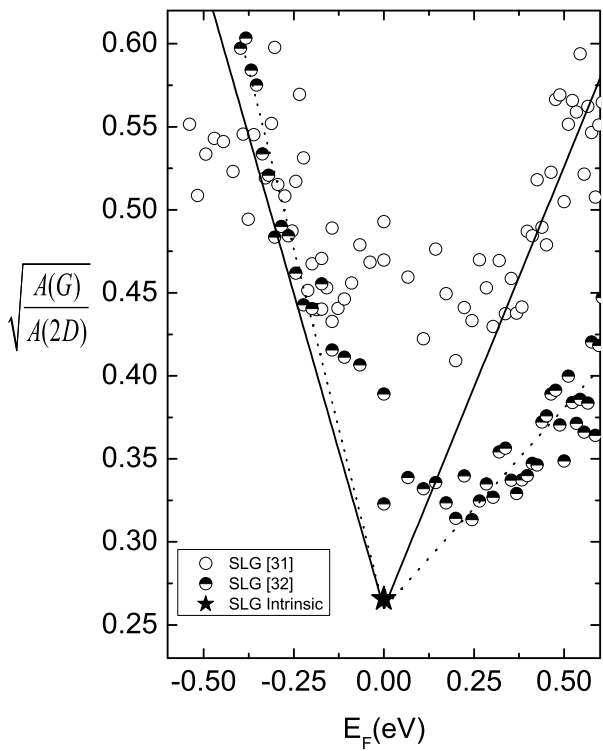}
\caption{Fit of the experimental dependence $\sqrt{A(G)/A(2D)}$
from Ref.~\onlinecite{DasCM} (open circles) and Ref.~\onlinecite{Dasbila}
(half-filled circles) using Eq.~(\ref{fitexpr2}) (dashed and solid lines,
respectively).}
\label{fig:fit}
\end{figure}
Fig.~\ref{fig:fit} plots $\sqrt{A(G)/A(2D)}$ as a function of $E_F$. This dependence, according to Eq.~(\ref{fitexpr2=}), should correspond to two symmetric straight lines joining at $E_F=0$. As noted in Sec.~I, close to $E_F=0$ the data from the two polymer electrolyte gating experiments do not converge to the same value. However, for both a linear rise of $\sqrt{A(G)/A(2D)}$ is seen at higher energies. Also, while the data represented by open circles in Fig.4 are almost symmetric, a significant asymmetry is seen for electron doping in the set represented by the half-filled circles, while the two sets are in good agreement for hole doping.

A(2D)/A(G) for intrinsic samples measured at 514.5 nm excitation, the same used in Refs.\onlinecite{DasCM,Dasbila}, is in the range 12-17\cite{ACFRaman,heinz,berciaud}, represented by the star in Fig. 4 at 14.5. This is in good agreement with the ratio measured for carbon whiskers\cite{tan1}. These show a 2D peak very similar to graphene, being composed of mis-oriented graphene layers\cite{tan1,latil}. However, their Raman spectra are much less susceptible to charged impurities or surface doping, being bulk materials\cite{tan1}. This corresponds to $\sqrt{A(G)/A(2D)}\sim0.24-0.29$, which we use to eliminate the effect of doping inhomogeneity, by constraining $\sqrt{A(G)/A(2D)}\sim0.26$ at zero doping. We also need to consider the dielectric constant of the polymer electrolyte~\cite{DasCM}, $\varepsilon=5$, giving $f(e^2/\varepsilon{v}_F)\approx{0}.06$. Thus, we fit the data with a one-parameter expression:
\begin{equation}\label{fitexpr2}
\sqrt{\frac{A(G)}{A(2D)}}=\frac{0.26}{\gamma_{e-ph}}\,
(\gamma_{e-ph}+0.06|\epsilon_F|).
\end{equation}
We fit separately each branch of the two data-sets, as shown by solid and dotted lines in Fig.~\ref{fig:fit}. We get $\gamma_{e-ph}$: 18, 21, 29, 65 meV, with an average $\gamma_{e-ph}\sim 33$~meV.
\section{Raman Intensities and Electron-Phonon Coupling}
\subsection{Theoretical Background and Electron-Phonon Coupling Definitions}
Even though graphite and other $sp^2$-hybridized materials have been investigated for more than 50 years\cite{Wallace1947,tuinstra}, all the fundamental physical properties needed for the interpretation of the Raman spectra have undergone an intense debate, which seems to be just beginning to converge. Interestingly, several features of both phonon dispersions and band structure of graphene are determined by the EPC. For example, in the Kohn anomalies around $\bf\Gamma$ or \textbf{K}\cite{PiscanecPRL} the correction to the phonon frequencies due to EPC results in a linear slope of the optical phonon branches as the wave vector approaches $\bf\Gamma$ or \textbf{K}. The EPC and phonon dispersions calculations of Ref.~\onlinecite{PiscanecPRL} have been confirmed at the $\bf \Gamma$ point by inelastic X-ray scattering\cite{Maultzsch2004}, and by the measurement of FWHM(G) in graphite, graphene and nanotubes\cite{LazPRB2006,piscaPRB,ACFRaman,Pisana}, once an-harmonic effects are taken into account\cite{Bonini2007,ACFRaman,Pisana}. For the \textbf{K} point, the precise slope of the anomaly is debated \cite{Lazzeri2008,Graf2007,Gruneis2009}. Another EPC effect is the kink in the electron dispersion, $\sim$200~meV below $E_F$, seen by angle-resolved photoemission spectroscopy (ARPES)\cite{Rotenberg2007,Zhou2008}. This is attributed to a correction to the electron energy due to EPC\cite{Zhou2006,Rotenberg2007,Zhou2008}, although alternative explanations also exist\cite{Olevano}. Thus, a correct EPC determination is a fundamental step for an accurate description of the physical properties of graphene, and nanotubes, being rolled up graphene sheets.

To link the 2D intensity to the EPC we first consider the rate of phonon emission by the
photoexcited electron/hole, $2\gamma_\mathrm{e-ph}$. This is obtained
from the imaginary part of the electron self-energy,
$\gamma_\mathrm{e-ph}=\Im\Sigma_\mathrm{e-ph}(\epsilon)$. For $E{_L}/2>E_F+\omega_\Gamma$, as in the case of the Raman measurements at 2.41 eV excitation of Refs.\onlinecite{DasCM,Dasbila}, we have\cite{BaskoBig}:
\begin{eqnarray}\label{gamma1=}
&&\gamma_K=\frac{\lambda_K}4\left(\frac{E{_L}}2-\omega_K\right),\quad
\gamma_\Gamma=\frac{\lambda_\Gamma}{4}\left(\frac{E{_L}}2-\omega_\Gamma\right)\nonumber\\
\end{eqnarray}
Then, from Eq.~(2):
\begin{equation}\label{gamma=}
\gamma_\mathrm{e-ph}
=\frac{\lambda_K}4\left(\frac{E{_L}}2-\omega_K\right)
+\frac{\lambda_\Gamma}{4}\left(\frac{E{_L}}2-\omega_\Gamma\right),
\end{equation}
The dimensionless coupling constants $\lambda_{\Gamma},\lambda_K$ correspond to phonons close to $\bf\Gamma$ and \textbf{K}, respectively, and determine their rate of emission. We define them as:
\begin{equation}\label{lambda=}
\lambda_{\Gamma,K}=\frac{F_{\Gamma,K}^2A_\mathrm{u.c.}}{2M\omega_{\Gamma,K}v_F^2}.
\end{equation}
Here $\omega_K=1210\:\mathrm{cm}^{-1}=0.150\:\mathrm{eV}$\cite{Gruneis2009} and $\omega_{\Gamma}=1580\:\mathrm{cm}^{-1}=0.196\:\mathrm{eV}$,\cite{ACFRaman} $M\approx{2}.00\cdot{10}^{-23}\:\mathrm{g}=2.88\cdot{10}^3\:(\mathrm{eV}\cdot\mbox{\AA}^2)^{-1}$ is the mass of the carbon atom, $A_\mathrm{u.c.}\approx{5}.24\:\mbox{\AA}^2$ is the unit cell area. $F_\Gamma$~and $F_K$ have the dimensionality of a force and are the proportionality coefficients between the change in effective hamiltonian and the lattice displacement along the corresponding phonon mode. Strictly speaking, the relevant phonon states are not exactly at ${\bf\Gamma}$ and \textbf{K}, as shown in Fig. 1. However, the corresponding deviation, $q\sim{E}_L/v_F$, is small compared to the \textbf{K}-\textbf{K'} distance, and is neglected. All observables depend on the dimensionless EPCs, $\lambda_{\Gamma}$ and $\lambda_K$.

Eq.~(\ref{lambda=}) follows the notation of Ref.~\onlinecite{BaskoBig}. Since different EPC definitions are used in the literature, it is quite useful to give here matching rules for all of them, which will be necessary when comparing the EPC values obtained here with previous (and future) reports. The EPCs can be conveniently matched by either relating them to the nearest-neighbor tight-binding model, where the constants are expressed in terms of a single parameter: $\partial{t}_0/\partial{a}$, the derivative of the nearest-neighbor electronic matrix element with respect to the interatomic distance, or by comparing expressions for various observables. For example, doping leads to a G peak shift due to EPC. This is expressed in terms of $E_F$ as\cite{Pisana,YanPrl2007,Lazzeri2006}:
\begin{equation}\label{YanEq=}
\delta\omega_\Gamma=\frac{\lambda_\Gamma}{2\pi}
\left(|E_F|+\frac{\omega_\Gamma}4
\ln\frac{2E_F-\omega_\Gamma}{2E_F+\omega_\Gamma}\right).
\end{equation}
The corrections to the phonon dispersions as function of wavevector~\textbf{q}, measured from $\bf\Gamma$ or \textbf{K}, are\cite{PiscanecPRL,piscaPRB,BaskoBig}:
\begin{subequations}\begin{eqnarray}
&&\delta\omega_{\Gamma-LO}=
\frac{\lambda_\Gamma}8\,\sqrt{v_F^2q^2-\omega_\Gamma^2},\label{wLO=}\\
&&\delta\omega_{\Gamma-TO}=
-\frac{\lambda_\Gamma}8\,\frac{\omega_\Gamma^2}{\sqrt{v_F^2q^2-\omega_\Gamma^2}},\\
&&\delta\omega_K=\frac{\lambda_K}4\,\sqrt{v_F^2q^2-\omega_K^2}.\label{omegaK=}
\end{eqnarray}\end{subequations}
Note that the $E_{2g}$ mode splits into longitudinal ($\Gamma-LO$) and transverse ($\Gamma-TO$) at finite~$q$. Note also that due to analytical properties of the logarithm and square root, Eq.~(\ref{YanEq=}) at $|E_F|<\omega_\Gamma/2$ and Eqs.~(\ref{wLO=})--(\ref{omegaK=}) at $v_Fq<\omega_{K,\Gamma}$ acquire imaginary parts, which correspond to the phonon decaying into a continuum of electron-hole pairs\cite{LazPRB2006}. In this case $2\Im\delta\omega$ gives the FWHM of the corresponding Lorentzian profile. At $v_Fq\gg\omega_{K,\Gamma}$ Eqs.~(\ref{wLO=}) and~(\ref{omegaK=}) give the profile of the Kohn anomalies.

In Refs.~\onlinecite{PiscanecPRL,Pisana,Calandra2007,Lazzeri2008} the EPCs are defined as the matrix elements of the Kohn-Sham potential,
differentiated with respect to the phonon displacements. What enters the observables are their squares, averaged over the Fermi surface in the limit $E_F\to{0}$. The matching rule is then:
\begin{subequations}\begin{eqnarray}
&&F^2_\Gamma=
4\langle D^2_{\bf \Gamma}\rangle_F^{\mbox{\scriptsize{(Refs.~\onlinecite{Pisana,Lazzeri2008})}}}
=8M\omega_\Gamma\langle g^2_{\bf \Gamma}
\rangle_F^{\mbox{\scriptsize{(Ref.~\onlinecite{PiscanecPRL,Calandra2007})}}}\nonumber\\
\label{D2Gamma=}\\
&&F^2_K=
2\langle D^2_{\bf K}\rangle_F^{\mbox{\scriptsize{(Refs.~\onlinecite{Pisana,Lazzeri2008})}}}
=4M\omega_K\langle g^2_{\bf K}
\rangle_F^{\mbox{\scriptsize{(Refs.~\onlinecite{PiscanecPRL,Calandra2007})}}}\nonumber\\
\end{eqnarray}\end{subequations}
In Ref.~\onlinecite{YanPrl2007} the dimensionless coupling constant~$\lambda$
is defined as the proportionality coefficient in Eq.~(\ref{YanEq=}). Thus,
\begin{equation}\label{lambdaYan=}
\lambda^{\mbox{\scriptsize (Ref.\onlinecite{YanPrl2007})}}=\frac{\lambda_\Gamma}{2\pi}.
\end{equation}
Note that the expression linking EPC to FWHM(G) in Ref.~\onlinecite{YanPrl2007}
underestimates FWHM(G) by a factor~2.

The dimensionless EPC reported in the ARPES analysis of Refs.~\onlinecite{Rotenberg2007,Zhou2008,Rotenberg2008,GruneisARPES} and in the scanning tunneling spectroscopy (STS) experiment of Ref.~\onlinecite{EvaAndrei} was measured from the ratio of the electronic velocities below and above the kink in the electron dispersion. This ratio is determined by the derivative of the real part of the electronic self-energy $\Re\Sigma_{e-ph}(\epsilon)$ due to the EPC. The latter can be calculated if one takes the Dirac spectrum for electrons and a constant dispersion for phonons. For $E_F>0$\cite{Calandra2007}:
\begin{eqnarray}
\Sigma_{e-ph}(\epsilon)&=&{}-\frac{\lambda_K}{4\pi}
(\epsilon-\omega_K)\ln\frac{E_M}{|\epsilon-\omega_K-E_F|}
\nonumber\\
&&{}-\frac{\lambda_K}{4\pi}(\epsilon+\omega_K)
\ln\frac{E_M|\epsilon+\omega_K-E_F|}{(\epsilon+\omega_K)^2}\nonumber\\
&&{}-\frac{\lambda_\Gamma}{4\pi}
(\epsilon-\omega_\Gamma)\ln\frac{E_M}{|\epsilon-\omega_\Gamma-E_F|}
\nonumber\\
&&{}-\frac{\lambda_\Gamma}{4\pi}(\epsilon+\omega_\Gamma)
\ln\frac{E_M|\epsilon+\omega_\Gamma-E_F|}{(\epsilon+\omega_\Gamma)^2}.
\end{eqnarray}
Here $E_M$ is the ultraviolet cutoff, of the order of the electronic bandwidth.
We then get the matching rule:
\begin{eqnarray}\nonumber
\lambda^\mathrm{(kink)}&=&
-\left.\frac{\partial\Re\Sigma_{e-ph}}{\partial\epsilon}
\right|_{\epsilon=E_F}\\
&=&\frac{\lambda_K}{2\pi}
\left(\frac{E_F-\omega_K}{\omega_K}
+\ln\frac{E_M}{\omega_K+E_F}\right)\nonumber\\
&&{}+\frac{\lambda_\Gamma}{2\pi}
\left(\frac{E_F-\omega_\Gamma}{\omega_\Gamma}
+\ln\frac{E_M}{\omega_\Gamma+E_F}\right).
\label{lambdaARPES=}
\end{eqnarray}
However, we note that $\lambda_K$ is subject to Coulomb renormalizations\cite{BaskoAleiner}. This implies that $\lambda_K$ depends on the electronic energy scale, such as the electron energy~$\epsilon$, the Fermi energy~$E_F$, or the temperature~$T$, whichever is larger: $\lambda_K=\lambda_K(\max\{|\epsilon|,|E_F|,T\})$. This dependence is shown in Fig.~6 of Ref.~\onlinecite{BaskoAleiner} in the semi-logarithmic scale. In a Raman measurement this scale is given by the energy of the photo-excited electron: $\epsilon\approx{E}_L/2$, as long as $E_L/2>|E_F|$. Thus, in Eq.~(\ref{gamma=}) $\lambda_K=\lambda_K(E_L/2)$. On the other hand, to estimate the EPC effects on the phonon dispersions in the intrinsic graphene, the relevant electron energy is of the order of the phonon energy. Thus, in Eq.~(\ref{omegaK=}) $\lambda_K \sim \lambda_K(\omega_K)$. From Fig.~6 of Ref.~\onlinecite{BaskoAleiner} we estimate that $\lambda_K(\omega_K)/\lambda_K(E_L/2) \approx{1}.5$ for $\varepsilon=1$ and 1.2 for $\varepsilon=5$ (taking $E_L\approx{2}$~eV to represent Raman measurements in the visible range).

The situation with Eq.~(\ref{lambdaARPES=}) is more complicated, since the cutoff~$E_M$ appears explicitly. The logarithmic term is determined by \emph{all} energy scales from $E_M$ down to $E_F+\omega_K$. Thus, the proper expression is
\begin{eqnarray}
\lambda^\mathrm{(kink)}&=&\frac{\lambda_K(E_F)}{2\pi}
\frac{E_F-\omega_K}{\omega_K}
+\int\limits_{E_F+\omega_K}^{E_M}
\frac{\lambda_K(\epsilon)}{2\pi}\,\frac{d\epsilon}{\epsilon}
\nonumber\\
&&+\frac{\lambda_\Gamma}{2\pi}
\left(\frac{E_F-\omega_\Gamma}{\omega_\Gamma}
+\ln\frac{E_M}{E_F+\omega_\Gamma}\right).\label{lambdaARPESint=}
\end{eqnarray}
\subsection{Experimental Electron-Phonon Coupling}
From Eq.(12), our overall average $\gamma_{e-ph}=33\:\mbox{meV}$, derived from a fit to all the data in Fig. 4, gives:
\begin{equation}\label{estimate=}
\lambda_\Gamma+\lambda_K\approx 0.13.
\end{equation}
On the other hand, the hole doping side of Fig. 4 shows two data sets very consistent with each other. We can thus get another estimate taken from the average $\gamma_{e-ph} \approx 20\:\mbox{meV}$ for just the hole doping side. This would give:
\begin{equation}\label{estimate2}
\lambda_\Gamma+\lambda_K\approx 0.08.
\end{equation}

Based on measurements\cite{Pisana,YanPrl2007} and DFT calculations\cite{PiscanecPRL}, the value of $\lambda_\Gamma$ can be reliably taken $\approx{0}.03$. Indeed, DFT gives~\cite{PiscanecPRL} $\langle g^2_{\bf \Gamma}\rangle_F=0.0405\:\mbox{eV}^2$ and $v_F=5.5\:\mbox{eV}\cdot\mbox{\AA}$, corresponding, from Eqs.~(\ref{lambda=}), (\ref{D2Gamma=}) to $\lambda_\Gamma\approx{0}.028$. Even though $\langle g^2_{\bf \Gamma}\rangle_F$ and $v_F$ are subject to Coulomb renormalization, $\lambda_\Gamma=4A_\mathrm{u.c.}\langle g^2_{\bf \Gamma}\rangle_F/v_F^2$, which contains their ratio, is not.\cite{BaskoAleiner} The experimental $\lambda_\Gamma$ extracted from FWHM(G) in graphene and graphite\cite{ACFRaman,LazPRB2006} according to Eq.~(\ref{wLO=}) and from the dependence of Pos(G) on Fermi energy according to Eq.~(\ref{YanEq=}), give $\lambda_{\Gamma}\approx{0}.034$\cite{YanPrl2007} and
$\lambda_{\Gamma}\approx{0}.027$\cite{Pisana}.

On the other hand, the value of $\lambda_K$ is still debated\cite{Calandra2007,BaskoAleiner,Lazzeri2008}. The calculated DFT $\langle g^2_{\bf K}\rangle_F=0.0994\:\mbox{eV}^2$, together with the DFT $v_F=5.5\:\mbox{eV}\cdot\mbox{\AA}$ (both taken from Ref.~\onlinecite{PiscanecPRL}) gives $\lambda_K=0.034$. However, Ref.~\onlinecite{BaskoAleiner} suggested this should be enhanced by Coulomb renormalization by up to a factor 3, depending on the background dielectric constant. In order to compare with our fits, we need consider that the corrections to the phonon dispersion are determined by electronic states with energies lower than those contributing to the Raman signal. As discussed in Sec.~IIIA, $\lambda_K(\omega_K)/\lambda_K(E_L/2) \approx 1.2$ for $\varepsilon=5$. Our fit in Eq.~(\ref{estimate=}) corresponds to $\lambda_K(E_L/2)\approx{0}.1$, while Eq.~(\ref{estimate2}) gives $\lambda_K(E_L/2)\approx{0}.05$, resulting in $\lambda_K(\omega_K)\approx{0}.12$ and $\lambda_K(\omega_K)\approx{0}.06$, respectively. These are bigger than DFT by a factor of about 3.5 and 1.7, respectively.

A recent GW calculation gave $\langle D^2_\mathbf{K}\rangle_F=193\:\mbox{eV}^2/\mbox{\AA}^2$\cite{Lazzeri2008}. Combining this with the GW $v_F=6.6\:\mbox{eV}\cdot\mbox{\AA}$\cite{Gruneis2008}, we get $\lambda_K(\omega_K)\approx{0}.054$, a factor $\sim 1.6$ greater than DFT, in good agreement with our fitted average on the hole side.

Ref.~\onlinecite{Gruneis2009} reported inelastic x-ray scattering measurements of the phonon dispersions near \textbf{K} more detailed than those originally done in Ref.~\onlinecite{Maultzsch2004}, now giving a phonon slope at~\textbf{K} of ${73}\:\mbox{meV}\cdot\mbox{\AA}$. Using Eq.~(\ref{omegaK=}) at $q\gg\omega_K/v_F$ and taking the experimental value $v_F=6.6\:\mbox{eV}\cdot\mbox{\AA}$\cite{Zhou2008} (the bare electron velocity, i.~e. below the phonon kink), we obtain $\lambda_K(\omega_K)\approx{0}.044$, a factor $\sim1.3$ higher than DFT, again in good agreement with our fitted average on the hole side.

Another EPC estimate can be derived from the 2D and 2D' area ratio. Combining Eqs.~(\ref{I2D=}),(\ref{I2Dp=}),12,13 we get:
\begin{equation}
\frac{A(2D)}{A(2D')}=2\left(\frac{\lambda_K}{\lambda_\Gamma}\right)^2
\end{equation}
For intrinsic SLG and graphite whiskers, the experimental A(2D)/A(2D') is $\sim25-30$~\cite{ACFRaman,heinz,berciaud,tan1}, which gives $\lambda_K(E_L/2)\approx{0}.11$ and $\lambda_\Gamma+\lambda_K(E_L/2)\approx0.13$. Since in this case $\varepsilon=1$, this results in $\lambda_K(\omega_K)\approx{0}.16$, a factor $\sim$4.5 higher than DFT, in agreement with our upper estimate from Eq. (22).

We finally consider the EPC derived from ARPES and STS. For an estimate, we approximate the dependence $\lambda_K(\epsilon)$ as linear in $\ln\epsilon$. We take $\lambda_K(E_M)=(\omega_\Gamma/\omega_K)\lambda_\Gamma$, as given by DFT (assumed to be valid at high energies), and leave $\lambda_K(E_L/2\approx{1}\:\mbox{eV})$ as the only free parameter determining this linear dependence:
\begin{equation}\label{linearlambda=}
\lambda_K(\epsilon)=\frac{\omega_\Gamma}{\omega_K}\,\lambda_\Gamma
-\left[\frac{\omega_\Gamma}{\omega_K}\,\lambda_\Gamma-\lambda_K(E_L/2)\right]
\frac{\ln(E_M/\epsilon)}{\ln[E_M/(E_L/2)]}.
\end{equation}
Taking $E_F=0.4\:\mbox{eV}$\cite{Rotenberg2007,Zhou2008,EvaAndrei,GruneisARPES}, $E_M=10\:\mbox{eV}$, and substituting Eq.~(\ref{linearlambda=}) in Eq.~(\ref{lambdaARPESint=}), we get:
\begin{equation}\label{lambdaA=}
\lambda^\mathrm{(kink)}\approx0.7\,\lambda_\Gamma+0.6\,\lambda_K(E_L/2).
\end{equation}
Note that the dependence on the precise value of $E_M$ is weak: setting $E_M=5\:\mbox{eV}$ changes the first
coefficient to $0.5$, and the second (more important as it multiplies the larger coupling constant) varies only by 2\%.
The measurements in Refs.~\onlinecite{GruneisARPES,Rotenberg2007,EvaAndrei,Rotenberg2008,Zhou2008} gave $\lambda^\mathrm{(kink)}\approx 0.4,\,0.3,\,0.26,\,0.2,\,0.14$, respectively. The smallest of these values, $\lambda^\mathrm{(kink)}\approx 0.14$, from Eq.~(\ref{lambdaA=}) corresponds to $\lambda_\Gamma+\lambda_K(E_L/2)\approx{0}.23$, while the highest to $\lambda_\Gamma+\lambda_K(E_L/2)\approx{0}.66$. Even the smallest is almost twice our upper bound fit of Eq.~(\ref{estimate=}) and would imply an EPC renormalization of almost one order of magnitude. Resolution effects could play a role in this overestimation\cite{Calandra2007}.

Thus, our fits to the doping dependent Raman area ratios point to a significant renormalisation, by a factor 1.7-3.5, of the TO mode close to \textbf{K}, responsible for the Raman D and 2D peaks. Our lower bound estimate is consistent with recent GW calculations and phonon measurements, but our upper bound is much lower than the smallest estimate derived by ARPES.
\section{Conclusions}
We have shown that the 2D intensity dependence on doping can be explained considering the influence of electron-electron interactions on the total scattering rate of the photogenerated electrons (holes). We have given a simple formula linking 2D peak area to the Fermi level shift. Fitting this to the available experimental data we got an estimate for the EPC value of the TO phonons close to \textbf{K}, responsible for the Raman D and 2D peaks. This is larger than that from DFT calculations, due to renormalisation by Coulomb interactions. However, our fitted EPC is still significantly smaller than those reported in ARPES or STS experiments.
\section{Acknowledgments} We acknowledge A. Das, S. Berciaud, A. Bonetti, P.H. Tan for useful discussions. A.C.F. acknowledges funding from the Royal Society and the European Research Council grant NANOPOTS.

\end{document}